\documentclass[preprint2]{aastex631}

\usepackage{amsmath}
\usepackage{graphicx}
\usepackage{acronym}

\begin{document}

\newcommand{\psr}{J0031$-$5726}
\newcommand{\DMunit}[0]{pc\,cm$^{-3}$}
\newcommand{\RMunit}[0]{rad\,m$^{-2}$}
\renewcommand{\software}[1]{\texttt{#1}}

\newcommand{\todo}[1]{{\color{red}#1}}

\acrodef{rfi}[RFI]{radio frequency interference}
\acrodef{vcs}[VCS]{Voltage Capture System}
\acrodef{mwa}[MWA]{Murchison Widefield Array}
\acrodef{rrat}[RRAT]{rotating radio transient}
\acrodef{gleamx}[GLEAM-X]{GaLactic and Extragalactic All-sky MWA eXtended}
\acrodef{gpm}[GPM]{Galactic Plane Monitor}
\acrodef{snr}[S/N]{signal-to-noise}
\acrodef{dm}[DM]{dispersion measure}
\acrodef{rm}[RM]{rotation measure}
\acrodef{ism}[ISM]{interstellar medium}
\acrodef{pa}[PA]{polarization angle}
\acrodef{rvm}[RVM]{Rotating Vector Model}
\acrodef{rfm}[RFM]{radius-to-frequency mapping}
\acrodef{ar}[AR]{aberration and retardation}
\acrodef{tec}[TEC]{total electron content}

\title{Discovery of an RRAT-like pulsar via its single pulses in an MWA imaging survey} 

\author[0000-0001-6114-7469]{Samuel J. McSweeney}
\affiliation{International Centre for Radio Astronomy Research, Curtin University, Bentley, WA 6102, Australia}

\author{Jared Moseley}
\affiliation{International Centre for Radio Astronomy Research, Curtin University, Bentley, WA 6102, Australia}

\author[0000-0002-5119-4808]{Natasha Hurley-Walker}
\affiliation{International Centre for Radio Astronomy Research, Curtin University, Bentley, WA 6102, Australia}

\author[0000-0001-5561-1325]{Garvit Grover}
\affiliation{International Centre for Radio Astronomy Research, Curtin University, Bentley, WA 6102, Australia}

\author[0009-0003-0996-9176]{Csan\'{a}d Horv\'{a}th}
\affiliation{International Centre for Radio Astronomy Research, Curtin University, Bentley, WA 6102, Australia}

\author[0000-0002-2801-766X]{Timothy J. Galvin}
\affiliation{CSIRO Space \& Astronomy PO Box 1130, Bentley, WA 6102, Australia}
\affiliation{International Centre for Radio Astronomy Research, Curtin University, Bentley, WA 6102, Australia}

\author[0000-0001-8845-1225]{Bradley W. Meyers}
\affiliation{International Centre for Radio Astronomy Research, Curtin University, Bentley, WA 6102, Australia}

\author[0000-0001-7509-0117]{Chia Min Tan}
\affiliation{International Centre for Radio Astronomy Research, Curtin University, Bentley, WA 6102, Australia}

\begin{abstract}

We report the discovery of PSR \psr{} in the GaLactic and Extragalactic All-sky MWA eXtended imaging survey at a Galactic latitude of $b \approx -60^\circ$.
The pulsar exhibits both sporadic, extremely bright pulses reminiscent of rotating radio transients (RRATs) as well as a persistent, dimmer pulses.
The bright pulses tend to arrive at later rotation phases than their dimmer counterparts, and have dramatically varying polarization angle curves, such that the integrated profile appears almost completely depolarized down to the system noise level.
The rotation measure of individual pulses was found to sometimes vary by up to ${\sim}0.8\,$rad$/$m$^2$, but was otherwise generally consistent with its average (ionosphere-corrected) value of $10.0 \pm 0.1\,$rad$/$m$^2$.
We surmise that \psr{} may represent a class of pulsar that is intermediate between normal pulsars and RRATs.

\end{abstract}

\section{Introduction}

Pulsars are rapidly rotating neutron stars which are known to exhibit a wide variety of single pulse behaviors, including sub-pulse drifting \citep{1968Natur.220..231D}, nulling \citep{1970Natur.228...42B}, giant pulse emission \citep{1968Sci...162.1481S,2006ApJ...640..941K}, and microstructure \citep{1971ApJ...169..487H}.
Various subclasses of pulsars can be defined in terms of the presence or absence these phenomena, as well as the time scales involved.
For instance, the subclass of \acp{rrat} \citep{McLaughlin2006} are those which exist in its null state for much longer---up to weeks, months, and years---than its burst state, which may only last for a few rotations \citep{2019JApA...40...42K}.
The subclass of giant pulse emitters can be defined as pulsars whose brightest pulses follow a power law distribution \citep[although other definitions exist, cf.][]{2004IAUS..218..315J}.
There does not yet exist a comprehensive model of pulsar radio emission that can satisfactorily account for all observed single pulse phenomena.
Finding more pulsars with interesting and unique single-pulse behavior, and studying them in detail, remains an important exercise towards gaining a deeper understanding of the emission mechanism.

Most known pulsars were discovered using Fourier-domain methods which exploit the strict periodicity of the pulses \citep[e.g.][]{2001MNRAS.328...17M,2010MNRAS.409..619K}.
Such methods are naturally less sensitive to pulsars with intermittent or strongly fluctuating pulses, which can dilute the \ac{snr} by leaking signal power away from the harmonics of the fundamental spin period.
Famously, \acp{rrat} are so intermittent that their defining characteristic is that they are more easily detected via their single pulses rather than via a periodicity search \citep{McLaughlin2006}.
Nowadays, radio transient surveys often include single-pulse search techniques as a standard part of their data processing pipelines \citep{2021ApJ...922...43G,2023MNRAS.524.5132D,2024arXiv240603806G}.

A less standard but increasingly common approach is to look for pulsars in imaging data \citep[e.g.][]{2013PASA...30....6M,2023PASA...40....3S}.
Although periodicity searches for ordinary pulsars (i.e. with periods $P \lesssim 1\,$s) are not viable with typical correlator integration times, pulsars can be identified in images by means of their time-averaged total intensity; (dispersed) single pulses \citep[e.g.][]{2021ApJ...922...43G,2023arXiv230900845S}; steep spectra \citep[e.g.][]{2018MNRAS.475..942F}; polarization \citep[e.g.][]{2018MNRAS.478.2835L}; and variability, due to interplanetary \citep[e.g.][]{2019PASA...36....2M} or interstellar scintillation \citep{2019MNRAS.482.2484B}.

Recent imaging surveys conducted with the \ac{mwa} \citep{Tingay2013} have explicitly included searches for radio transients, including the \ac{gleamx} survey \citep{2022PASA...39...35H}, and the \ac{gpm} survey (Hurley-Walker et al. in prep).
These surveys have detected, respectively, three ultra-long period radio transients: GLEAM-X\,J162759.5$-$523504.3 \citep[period $P\sim18$\,min and pulse width $W\sim30$--$60$\,s;][]{2022Natur.601..526H}; GPM\,J1839$-$10 \citep[$P\sim22$\,min, $W\sim20$--$300$\,s;][]{2023Natur.619..487H}; and GLEAM-X\,J0704$-$37 \citep[$P\sim2.9$\,hr; $W\sim30$\,s][]{2024arXiv240815757H}.

In this paper, we report the discovery and characterization of PSR \psr{}, an interesting \ac{rrat}-like pulsar, in the second data release of \ac{gleamx} \citep[spanning $20\,\mathrm{h}< \mathrm{RA} < 08\,\mathrm{h}$ and $-90^\circ < \mathrm{Dec} < 30^\circ$;][]{2024arXiv240606921R}.
Initially suspected to be an \ac{rrat}, follow-up with archived voltage data revealed \psr{} to be a $P{\sim}1.57\,$s pulsar with low \ac{snr} regular pulses, occasionally punctuated with extraordinarily bright, narrow bursts reminiscent of the bursty behavior characteristic of \acp{rrat}.

Details of the initial discovery, localization, follow up observations, and preliminary timing analysis are given in \S\ref{sec:observations}.
A single pulse analysis is given in \S\ref{sec:pulse_analysis}, including their polarization behavior (\S\ref{sec:polarization}).
We discuss the classification of \psr{} and compare it with similar pulsars and \acp{rrat} in \S\ref{sec:discussion}.
A brief summary of our findings is given in \S\ref{sec:conclusions}.

\section{Discovery, localization, and follow-up observations}
\label{sec:observations}

\subsection{Discovery}
\label{sec:discovery}

\psr{} was discovered in \ac{mwa} correlator observation 1286031336 (2020-10-06 14:55:18 UTC, MJD 59128.6217) as part of the \ac{gleamx} survey under project code G0008 \citep[][see Table \ref{tbl:observations}]{2022PASA...39...35H}, with a per-observation instantaneous bandwidth of 30.72~MHz.
The correlated visibilities have a time/frequency resolution of $0.5\,$s/$40\,$kHz, where each ``snapshot'' observation is a ${\sim}2$-minute drift scan with the primary beam centered on the local meridian, or $\pm1$\,hour either side.
\ac{gleamx}'s primary data product are images formed by integrating the visibilities over the entire 2-minute snapshot and mosaicked together \citep[available as part of Data Release II;][]{2024arXiv240606921R}.

To search for transients, a step is included in which the imaged model is subtracted, and a (dirty) image made of each timestep.
This results in a data cube in which a time series is associated with each pixel of the image.
These time series are then passed through a selection of filters designed to be sensitive towards different types of transient events.
A full description of the filters, the search, and the results will be described in full at a later date (Horv\'{a}th et al. in prep).
\psr{} was detected as a source which appeared as a series of unresolved (i.e. $W<4$\,s) bursts with peak flux density $S_\mathrm{154\,MHz}\sim$2\,Jy and no obvious periodicity.
Its low \ac{dm} of $\sim6$\,\DMunit{} (whose relative delay of ${\sim}1\,$timestep across the instantaneous \ac{mwa} band was barely visible in dynamic spectra formed from the correlator data) indicated that it was a Galactic source, likely an \ac{rrat} or pulsar.

\subsection{Localization}
\label{sec:localization}

Two particularly bright bursts were detected 34.75 and 61.25 seconds, respectively, into the discovery observation.
By imaging the field-of-view for one second starting at each of these times across 139--170\,MHZ, we obtained an image with an r.m.s. noise of 32\,mJy\,beam$^{-1}$, in which \psr{} appears as a $\sim$1\,Jy source (i.e. with \ac{snr}$\sim$30).
The synthesised beam resolution was $97''\times69''$ at a position angle of $147^\circ$ from North.
\textsc{fits\_warp} \citep{2018A&C....25...94H} was used to correct for the refractive position shifts induced by the ionosphere, bringing the astrometry in line with the Sydney University Molonglo Sky Survey \citep[SUMSS;][]{2003MNRAS.342.1117M}.
The position of \psr{} was shifted by $14''$ in this process, to a final position of 00:31:31.8 $-$57:26:37, to an accuracy of $2''$.

\subsection{\acs{vcs} follow-up and preliminary timing}
\label{sec:vcs-follow-up}

The pulses detected in the images showed no obvious periodicity, but the small (but non-zero) dispersion was taken as evidence that the source may be a Galactic intermittent pulsar or \ac{rrat}.
To test this, we performed a targeted periodicity search in the \ac{vcs} observation taken in 2018 as part of the Southern-sky \ac{mwa} Rapid Two-metre (SMART) pulsar survey covering this part of the sky \citep[Obs ID 1224252736, designated `B09', MJD 58413.5916; see][]{2023PASA...40...21B}.
The \ac{vcs} data were calibrated using a solution derived from a dedicated calibration observation and reduced with \software{hyperdrive} (Jordan et al., submitted)\footnote{\url{https://mwatelescope.github.io/mwa_hyperdrive/}}, a GPU-accelerated re-implementation of the Real Time System \citep{Mitchell2008} and \software{calibrate} \citep{Offringa2014}.
The data were then beamformed towards \psr{} using the method described by \citet{Ord2019}.
The beamformed data were written out as full Stokes samples with a time/frequency resolution of $100\,\mu$s/$10\,$kHz in the PSRFITS format \citep{Hotan2004}.

\begin{figure}[th]
    \centering
    \includegraphics[width=0.9\linewidth]{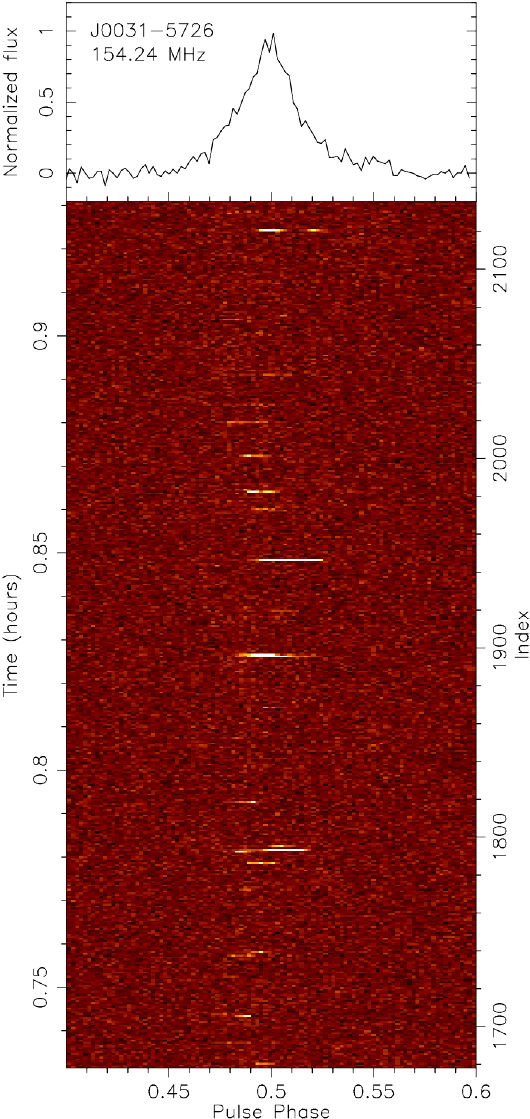}
    \caption{A portion of the pulse stack (bottom panel) from the 2018-10-22 observation of \psr{}, with ${\sim}3\,$ms time resolution. The profile (top panel) is formed from the whole ${\sim}80\,$min observation, not just the portion shown in the pulsestack. The phase has been manually aligned such that the pulse phase of 0.5 occurs at the peak of the profile. In order to make some of the relatively dim pulses visible, the color scale was saturated to 20\% of the brightest pixel value.}
    \label{fig:12242pav}
\end{figure}

Using \software{PRESTO} \citep{Ransom_New_search_techniques_2001,2011ascl.soft07017R}, a repeating source was detected at the source's coordinates with a barycentric period of ${\sim}1.57$\,s and a \ac{dm} of $6.73$\,\DMunit, confirming that the source is indeed a pulsar.
The frequency-scrunched pulse stack of this observation, a portion of which is shown in Fig.~\ref{fig:12242pav}, reveals the appearance of several exceptionally bright single pulses irregularly spaced throughout the time series, supporting the view that this is an \ac{rrat}.
The bright pulses appear to have an approximate wait time of a few minutes, with the intermediate pulses being either nulls or too dim to be seen above the noise floor.
We later show (in \S\ref{sec:pulse_analysis}), however, that there is faint emission even in (at least some of) the apparent nulls.

\begin{figure}
    \centering
    \includegraphics[width=\linewidth]{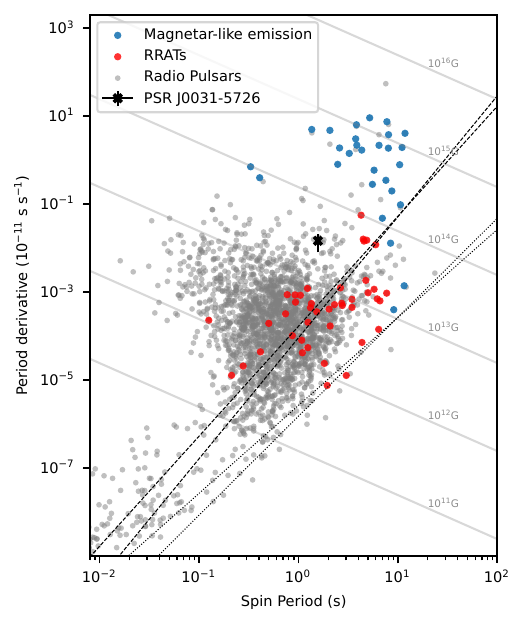}
    \caption{Spin period against spin period derivative for all known pulsars \citep{2005AJ....129.1993M} with the addition of the newly discovered \psr{}. The known pulsars are reported by the Australia Telescope National Facility, catalog version 2.0.1 \citep{2005AJ....129.1993M}. The dashed lines correspond to the theoretical death lines for a pure dipole and the dotted lines for twisted dipole \citep{2000ApJ...531L.135Z,1993ApJ...402..264C}. The measured $\dot{P}$ of \psr{} might be inaccurate, as discussed in the main text. If $\dot{P}$ turns out to be overestimated, it is possible that \psr{} sits closer to the main distribution of \acp{rrat}.}
    \label{fig:PPdot}
\end{figure}

The detection observation is a correlator observation with an integration time of $0.5\,$s, which is too coarse for furthering the timing analysis.
However, a search\footnote{\url{https://github.com/cplee1/mwa_source_finder}} in the MWA archives revealed a 2016 \ac{vcs} observation (Obs ID 1163853320, MJD 57714.5243) with the source in the field of view for long enough to obtain a second measurement of the period, $P$.
For both observations, we used the PRESTO-derived period and \ac{dm} to dedisperse the data and fold them with \software{DSPSR} to create 10-second, frequency-scrunched archives with 1024 phase bins.
The barycentric period for each observation was obtained using the \software{PSRCHIVE} utility \software{pdmp}, which are given (along with other details of the observations) in Table \ref{tbl:observations}.
The measured periods imply a spindown rate of $\dot{P} = (1.5 \pm 0.7) \times 10^{-13}\,$s$/$s between 2016 and 2018.
The location of \psr{} on the $P$-$\dot{P}$ diagram is given in Fig.~\ref{fig:PPdot}.

We note, however, that the correlation between pulse energy and pulse phase (described in \S\ref{sec:pulse_analysis}) may introduce a systematic error into the period measurements, which may in turn affect $\dot{P}$.
A proper timing campaign would be needed to obtain a robust measurement of $\dot{P}$.

With the preliminary timing solution in hand, the two \ac{vcs} observations were folded to make single-pulse archives using \software{DSPSR} \citep{VanStraten2011b}.
The \ac{snr} of the 2016 observation, while adequate for a period measurement, was deemed too low to be useful for the single-pulse analysis described in \S\ref{sec:pulse_analysis}, which we therefore only undertook for the 2018 observation.
A small amount of \ac{rfi} was found in the 2018 observation, identified by manual inspection of the pulse stacks and single pulse dynamic spectra.
As a result of manual flagging, 137 pulses were discarded (out of 3051) in which relatively short duration \ac{rfi} was seen to contaminate most of the band.
In most other cases of \ac{rfi} (${\sim}130$ pulses), we flagged only a small number of affected channels (never more than $1.3\,$MHz out of $30.72\,$MHz).
A portion of the pulse stack of the 2018 observation shown in Fig.~\ref{fig:12242pav} is after \ac{rfi} flagging.

\begin{deluxetable*}{cccccccc}
    \tablecaption{\acs{mwa} observations of \psr{}. All observations have a total bandwidth of 30.72~MHz. \label{tbl:observations}}
    \tablehead{
        \colhead{Obs ID} & \colhead{Year} & \colhead{MJD} & \colhead{Duration} & \colhead{Frequency} & \colhead{Cal ID} & \colhead{$P$ (\software{pdmp})} & \colhead{Obs. type} \\
        & & & \colhead{(s)} & \colhead{(MHz)} & & \colhead{(s)} &
    }
    \startdata
    1163853320 & 2016 & 57714.5243 & 5120  & 184.96 & 1163848248 & 1.570319(4) & \ac{vcs} \\
    1224252736 & 2018 & 58413.5916 & 4792 & 154.24 & 1224277624 & 1.570328(1) & \ac{vcs} \\
    1286031336 & 2020 & 59128.6217 & 120 & 154.24 & (self-cal.) & - & Correlator \\
    \enddata
\end{deluxetable*}

\subsection{Polarization}

The \ac{vcs} observations are dual-polarization, and the standard processing described above generates full Stokes pulsar archives.
The calibration used only a Stokes I sky model, making it possible for the unconstrained phase delay between the two orthogonal polarizations to generate leakage between the various Stokes parameters \citep{Ord2019}.
Inaccuracies in the assumed beam model \citep{2017PASA...34...62S} can cause similar leakage.

\begin{figure}
    \centering
    \includegraphics[width=\linewidth]{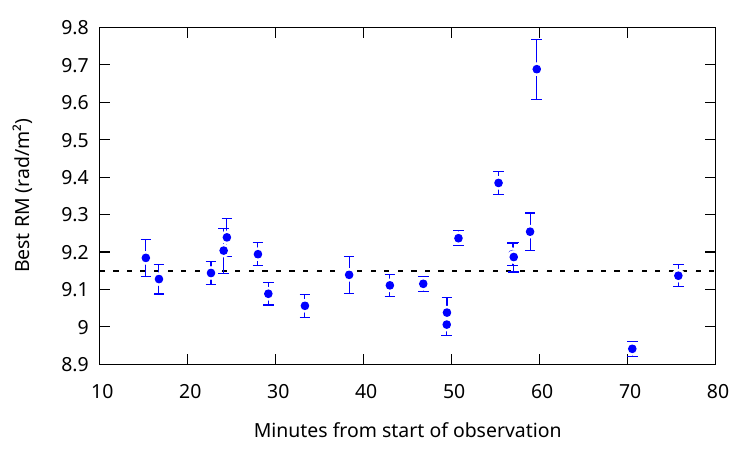}
    \caption{\ac{rm} measured for a selection of the brightest individual pulses. The dashed horizontal line is the weighted average, ${\rm RM}_{\rm meas} = 9.15 \pm 0.04\,$rad$/$m$^2$. The variability of \ac{rm} over the course of the observation is apparently real, and larger than seen in other pulsars. Whether it could be ionospheric in origin or intrinsic to the source is discussed in \S\ref{sec:polarization}.}
    \label{fig:RM_results}
\end{figure}

As we did not perform any explicit polarisation calibration, we used the pulsar observations themselves to estimate the leakage.
Pulsar signals are Faraday rotated by the \ac{ism} (as well as the Earth's ionosphere), so leakage from Stokes I and V into Q and/or U can be detected by excess power in the \ac{rm} power spectrum at 0\,rad$/$m$^2$, while leakage \emph{between} Q and U will appear at the opposite sign to the true \ac{rm}.
As the brightest pulses have a higher \ac{snr} than the total profile, we measured the \ac{rm} on individual pulses using \software{PSRSALSA} \citep{2016A&A...590A.109W}, shown in Fig.~\ref{fig:RM_results}.
The weighted average \ac{rm}, which includes both \ac{ism} and ionosphere contributions, is ${\rm RM}_{\rm meas} = 9.15 \pm 0.04$\,rad$/$m$^2$.
We did indeed see a small amount of excess power at 0~rad$/$m$^2$, but no excess at the opposite sign, estimating the leakage (from either Stokes I or V into Q and U) to be $\lesssim 10\%$.
While this could be improved with more careful calibration, it is sufficient to demonstrate \psr{}'s interesting single-pulse polarization characteristics, discussed in \S\ref{sec:polarization}.

Using \software{ionFR}\footnote{\url{https://github.com/csobey/ionFR}, git version \texttt{4fe4f12}}\citep{2013A&A...552A..58S}, we estimated the contribution of the ionosphere to the measured \ac{rm} at the time of the 2018-10-22 observation to be ${\rm RM}_{\rm ion} = -0.9 \pm 0.1\,$rad$/$m$^{2}$, and the contribution from the \ac{ism} to therefore be
\begin{equation}
  \begin{aligned}
    {\rm RM}_{\rm ISM}
      &= {\rm RM}_{\rm meas} - {\rm RM}_{\rm ion} \\
      &= 10.0 \pm 0.1 \, {\rm rad}/{\rm m}^2.
  \end{aligned}
\end{equation}

\psr{}'s basic properties, both measured and derived, are summarized in Table~\ref{tbl:ephemeris}.

\subsection{Summary of observations}

The three \ac{mwa} observations discussed in this paper are listed in Table \ref{tbl:observations}, but were introduced individually in different sections above.
For clarity, we briefly summarize each observation's role in the analysis presented in this paper.

The 2016 observation is a \ac{vcs} observation with a native time resolution of $100\,\mu$s. Its \ac{snr} was not high enough to warrant performing a single pulse analysis, but was high enough to obtain a measurement of the period, used as a lever arm for constraining \psr{}'s spindown rate.

The 2018 observation is also a \ac{vcs} observation with $100\,\mu$s time resolution, and, like the 2016 observation, was also used in the timing analysis.
It is the only observation in our set with both sufficient time resolution and high enough \ac{snr} to make it suitable for a detailed single-pulse analysis, detailed in the following sections.

The 2020 observation is a correlator observation with an integration time of 0.5 seconds, rendering it also unsuitable for single pulse analysis. Apart from being the discovery observation, this data set was used for imaging the field and localizing the source.

\begin{table*}
\centering
\caption{Properties of \psr{} derived from \acs{mwa} data\label{tbl:ephemeris}}
\begin{tabular}{ll}
    \hline\hline
    \multicolumn{2}{c}{Best-fit metrics and timing} \\
    \hline
    Pulsar name & \psr  \\
    Right Ascension (J2000) & 00:31:31.8(1) \\
    Declination (J2000) & $-$57:26:37(2) \\
    Spin frequency, $f$ (Hz) & 0.6368097(4) \\
    Reference epoch (MJD) & 58413.622117 \\
    First spin frequency time derivative, $\dot{f}$ (${\rm Hz}^2$) & $-6(3) \times 10^{-14}$ \\
    Dispersion measure, DM (\DMunit) & 6.755(32)  \\
    Rotation measure, RM (\RMunit) & 10.0(1) \\
    \hline
    \multicolumn{2}{c}{Derived quantities} \\
    \hline
    Galactic longitude, $l$ (deg) & 308.206 \\
    Galactic latitude, $b$ (deg) & $-$59.480 \\
    NE2001 distance (kpc) & 0.37 \\
    YMW16 distance (kpc) & 0.59 \\
    Spin period, $P$ (s) & 1.570328(1) \\
    First spin period time derivative, $\dot{P}$ (s$/$s) & $1.5(7) \times 10^{-13}$ \\
    Surface magnetic field strength, $B_{\rm s}$ (G) & $1.5 \times 10^{13}$ \\
    Spin-down luminosity, $\dot{E}$, ($\rm erg\, s^{-1}$) & $1.5 \times 10^{33}$ \\
    Characteristic age, $\tau_{\rm c}$ (Myr) & 0.17 \\
    \hline
\end{tabular}
\end{table*}

\section{Analysis and Results}
\label{sec:analysis}

\subsection{Single pulse statistics}
\label{sec:pulse_analysis}

The pulse stack in Fig.~\ref{fig:12242pav} shows the presence of several \ac{rrat}-like bright bursts that occur with a wait time of a few minutes.
However, some of the visible pulses are only just above the noise level, and flagging all pulses above the 1$\sigma$ level of the off-pulse noise and integrating the remaining pulses reveals that the apparent ``nulls'' still contain a significant and detectable pulsar signal.

To test whether any of the pulses with no apparent emission were truly nulls, we used Gaussian mixture modelling method\footnote{We used the implementation \texttt{pulse\_nulling}, \url{https://github.com/AkashA98/pulsar_nulling}, described more fully in \citet{Anumarlapudi2023}} described by \citet{Kaplan2018} and \citet{Anumarlapudi2023}.
This algorithm decomposes the pulse energy distribution into individual components, taking into account the known distribution of noise that can be measured from the off-pulse region.
It assumes the presence of a ``nulling'' component having the same distribution as the off-pulse noise, designed to avoid the bias introduced by low \ac{snr} pulses that would affect earlier methods such as \citet{Ritchings1976}.

\begin{figure}
    \centering
    \includegraphics[width=\columnwidth]{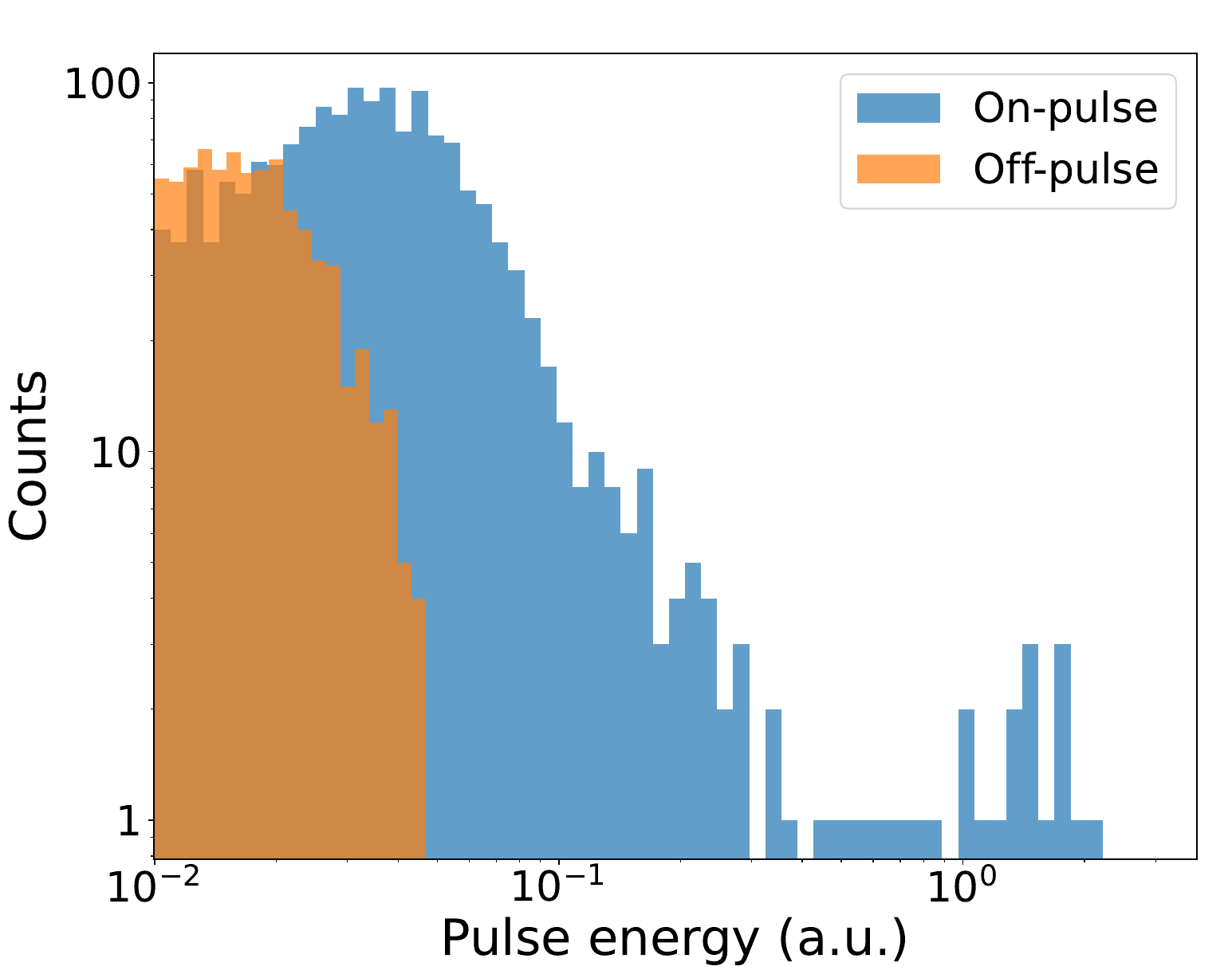}%
    \caption{The distribution of pulse energies (blue) in the 2018 \ac{vcs} observation, where the pulse energy is the integrated flux density over the estimated pulse window size of $0.08\,P$. The high energy tail is not well fit by a Gaussian, indicative of the inapplicability of the Gaussian mixture model to this data set, as discussed further in the main text. The off-pulse distribution (orange) was computed for an equal-sized phase range far away from the pulse window, dominated by system noise. Counts of both on-pulse and off-pulse energies smaller than $10^{-2}$ arbitrary units (including negative values due to noise fluctuations) are not shown on this plot.}
    \label{fig:nullfits}
\end{figure}

The pulse energies of the 2018 observation are shown in Fig.~\ref{fig:nullfits}, revealing a long-tailed distribution. The Gaussian mixture algorithm performed poorly for this data set, which we believe is partly because the distribution of bright pulses was not well fit with a Gaussian component, but also possibly because of the intrinsic difficulty of distinguishing between the distribution of ``true'' nulls from pulses with extremely low \ac{snr}.
We also found that the analysis was unusually sensitive to the range of phase bins selected as the ``on-pulse region'' \citep[cf.][]{2024ApJ...970...78G}.

We then used the \software{penergy} utility from \software{PSRSALSA} to measure individual pulse statistics, namely, the pulse phase, width, integrated energy, and \ac{snr}.
In its ``on-pulse, burst-all'' mode, \software{penergy} searches for pulses by convolving a user-specified on-pulse region with boxcars of every possible width (up to an integer number of phase bins).
The reported pulse phase, width, and integrated energy are derived directly from the best-fitting boxcar, and the \ac{snr} is the ``summed on-pulse intensity, divided by the expected noise resulting from adding the given number of on-pulse bins together, each containing a rms equal to that of the off-pulse region'' (\software{PSRSALSA} documentation\footnote{\url{http://www.jb.man.ac.uk/~wltvrede/psrsalsa.pdf}}, accessed 14 January 2025).
For this search, we used 2048 phase bins ($0.77\,$ms per bin), defined the on-pulse region to the phase range 0.44 to 0.57, and defined the off-pulse region as all phases outside of the on-pulse region range.
The analysis that follows uses the reported \ac{snr} instead of integrated energy, as we found that the former was better suited for characterising dim pulses dominated by noise.

\begin{figure}[!htbp]
    \centering
    \includegraphics[width=\linewidth]{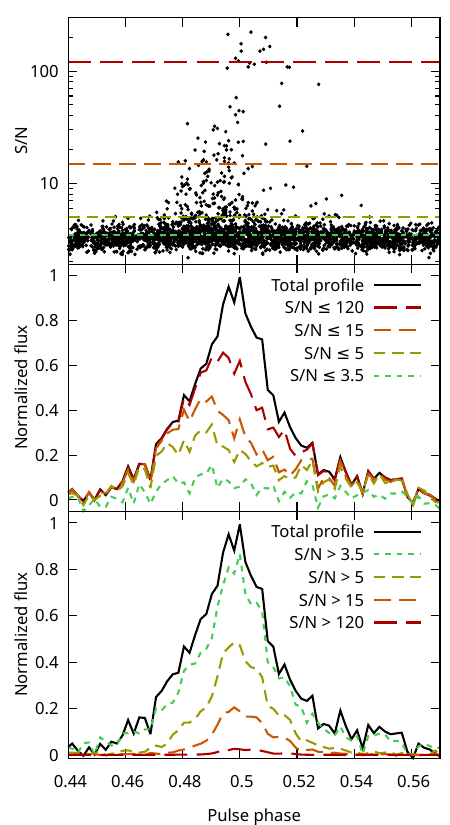}
    \caption{The top panel shows the individual pulse \acp{snr} as a function of pulse phase, as determined by \software{PSRSALSA}'s \software{penergy} utility. The horizontal lines drawn at \acp{snr} of 120, 15, 5, and 3.5 represent the maximal (middle panel) and minimal (bottom panel) thresholds used to form ``partial'' profiles normalized to the peak of the summed total profile (solid black lines), and binned to $3\,$ms resolution. The phase alignment is the same as that used in Fig.~\ref{fig:12242pav}.}
    \label{fig:snr_vs_phase}
\end{figure}

Fig.~\ref{fig:snr_vs_phase} shows the relation of \ac{snr} against pulse phase, revealing an interesting tendency for the very brightest pulses to lie preferentially on the trailing side of the pulse window, and the dimmer pulses to be clustered on the leading side.
To explore this further, we ordered the pulses by \ac{snr} and formed ``partial'' profiles by summing only pulses with \acp{snr} below a set of given thresholds, and normalizing them to the peak of the summed total profile.
The \ac{snr} thresholds (120, 15, 5, and 3.5) are somewhat arbitrarily chosen to make the profiles themselves easy to compare (especially the middle panel).
These thresholds correspond to including the brightest 10, 60, 224, and 1138 pulses (out of 2913) respectively.
The tendency for dimmer pulses to arrive earlier than the bright pulses is borne out in these profiles.
The weakest profile, formed from the dimmest ${\sim}63\%$ of the pulses, still has discernible pulsar signal in it.

The bottom panel of Fig.~\ref{fig:snr_vs_phase} shows the profiles formed by the pulses above each of the \ac{snr} thresholds defined above, normalized in the same way.
The peaks of these profiles remain much closer to phase 0.5, consistent with the expectation that the total profile is dominated by the brightest pulses.
These profiles also get narrower as fewer pulses are included, but this was found to be due to the distributions of single pulse arrival phases in each \ac{snr} range, and not, for example, to any correlation between pulse \ac{snr} and width (as discussed below).

\begin{figure}[!htbp]
    \centering
    \includegraphics[width=\linewidth]{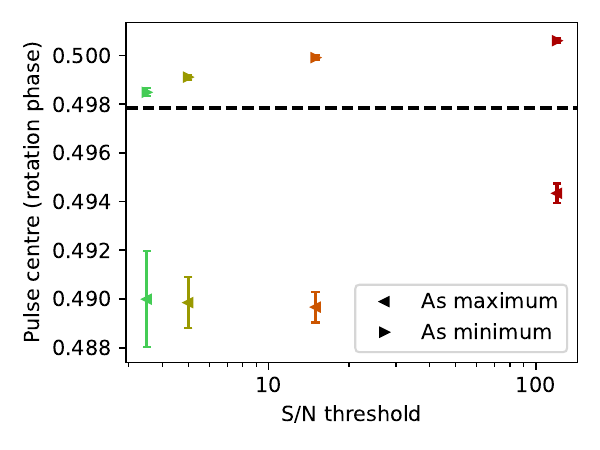}
    \caption{The fitted centers of Lorentzian fits to the partial profiles shown in Fig.~\ref{fig:snr_vs_phase} (with colors chosen to match). The ``As maximum'' points correspond to Fig.~\ref{fig:snr_vs_phase}'s middle panel, and the ``As minimum'' points correspond to its bottom panel. The horizontal, black dashed line indicates the fitted center of the total profile at rotation phase $0.4978 \pm 0.0003$.}
    \label{fig:profile_shift}
\end{figure}

Each partial profile was then fitted with a Lorentzian\footnote{The Lorentzian was preferred over the Gaussian as the fitted parameters had slightly (albeit marginally) lower uncertainties.} (using \software{curve\_fit} from the SciPy library) to determine how much the profile peak had shifted in each case.
These results are shown in Fig.~\ref{fig:profile_shift}.
Profiles dominated by the lower \ac{snr} pulses are centered around rotation phase 0.49, while those dominated by the higher \ac{snr} pulses approach, and even exceed, rotation phase 0.5.
The fitted center of the total profile is, as expected, intermediate between these two extremes, at around phase 0.498.

Further to this, we looked for other correlations between the other properties output by \software{PSRSALSA}.
We found no significant correlation between pulse width and either pulse phase, \ac{snr}, or integrated energy.

\subsection{Polarization}
\label{sec:polarization}

\begin{figure*}[p]
    \centering
    \includegraphics[width=\linewidth]{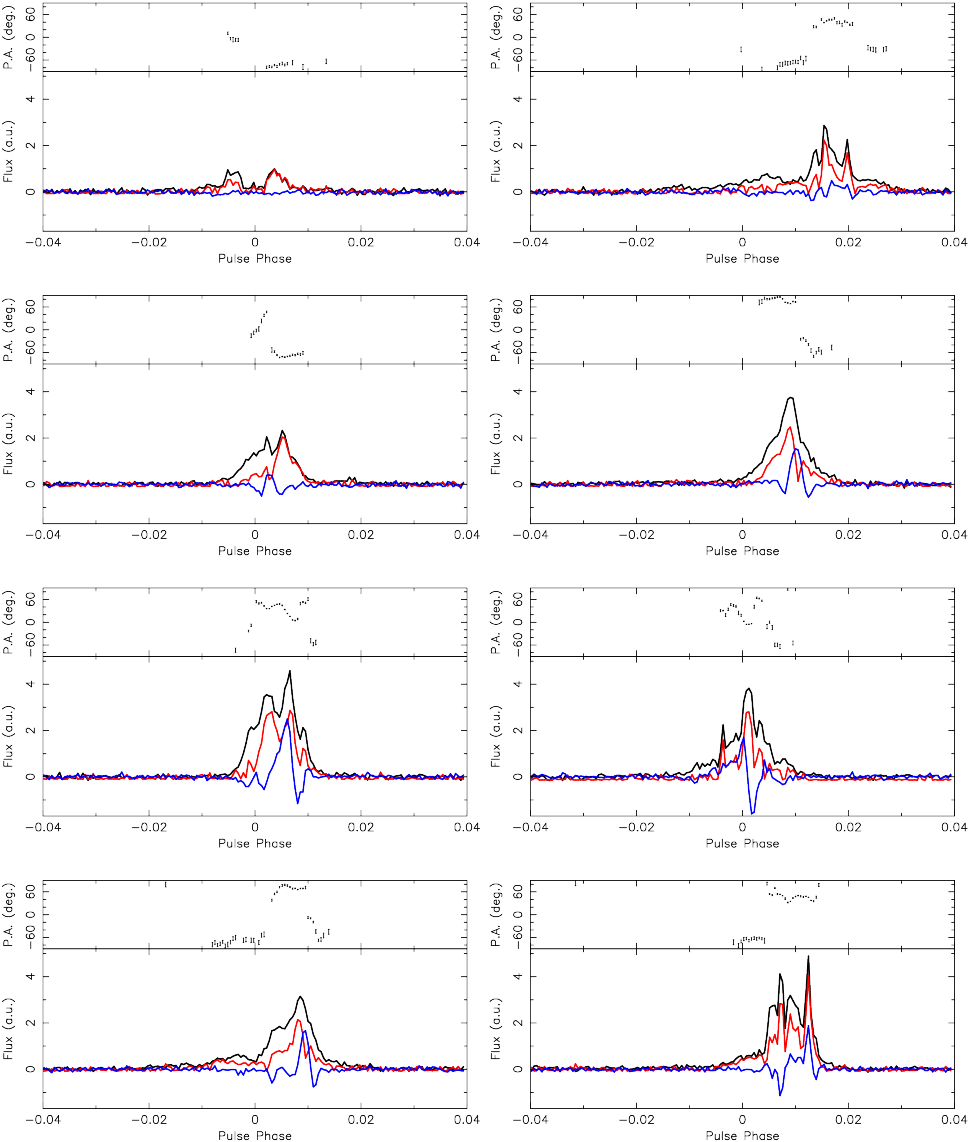}
    \caption{A selection of polarization profiles of some of the bright pulses in the 2018 observation. In each subplot, the polarization angle is shown in top panel, with the total intensity (black), linear polarization (red), and circular polarization (blue) shown in the bottom panel. The flux values are in arbitrary units, but all panels share the same the flux scale. The abscissa is shifted by $-0.5$ phase units relative to Figs.~\ref{fig:12242pav} and \ref{fig:snr_vs_phase}.}
    \label{fig:polprofs}
\end{figure*}

Polarization profiles of the single pulses from the 2018 observation were generated using \software{PSRCHIVE}, a selection from among the brightest of which is shown in Fig.~\ref{fig:polprofs}.
The \ac{pa} changes dramatically from pulse to pulse with no obvious pattern or predictable behavior.
The average profile shows negligible polarization, which can be understood by virtue of the fact that it is dominated by a relatively small number of the brightest pulses; their individual \acp{pa} vary so much that averaging them together results in a severely depolarized profile.

As can be seen in Fig.~\ref{fig:polprofs}, the \ac{pa} curves of the single pulses do not exhibit the characteristic ``S'' shape of the \ac{rvm} \citep{1969ApL.....3..225R}.
Instead, they exhibit sudden, unresolved jumps as well as slower variations on time scales on the order of a few milliseconds.
The jumps are often approximately 90 degrees, indicating the presence of distinct orthogonal modes seen in many other pulsars \citep{1969ApJ...158L...1E,1975ApJ...196...83M,2016ApJ...833...28M}.

\begin{figure}[th!]
    \centering
    \includegraphics[width=\linewidth]{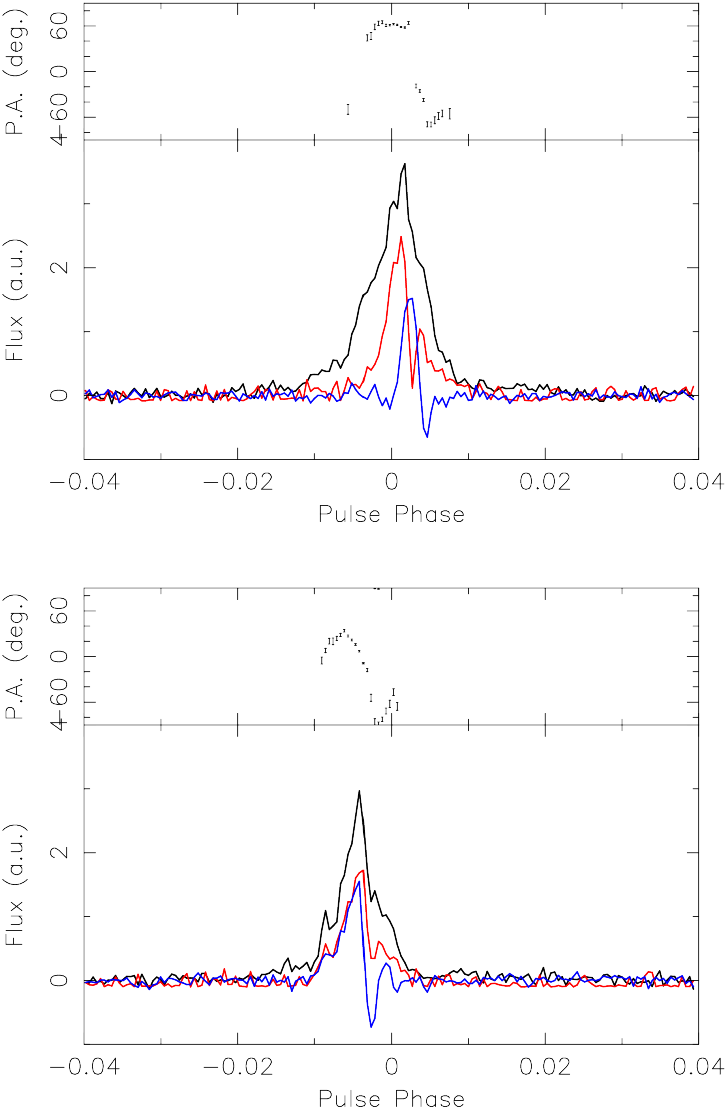}
    \caption{Same as Fig.~\ref{fig:polprofs} but showing two consecutive pulses, highlighting how the polarization behavior can change on a time scale of one rotation. The pulse shown in the top panel precedes the pulse in the bottom panel.}
    \label{fig:polprofs2}
\end{figure}

There is a single instance where two consecutive pulses were both bright enough to see their polarization in detail, shown in Fig.~\ref{fig:polprofs2}.
Here, the \ac{pa} curves have approximately the same overall shape, but shifted both in \ac{pa} and in rotation phase.
The circular polarization is also qualitatively similar, rising to a maximum fractional (V/I) value of a little over 50\%, and oscillating in handedness throughout the pulse.

As is evident in Fig.~\ref{fig:RM_results}, the measured RM also changes over the course of the observation, most notably from a maximum of ${\sim}9.69\,$rad$/$m$^2$ to a minimum of ${\sim}8.94\,$rad$/$m$^2$ in just over 10 minutes.
Geomagnetic storms have been observed in which small scale plasma disturbances can change the \ac{tec} by up to 10~TECU$/$min on 30-second time scales \citep[][see their Fig. 3]{1997GeoRL..24.2283P}, where 1~TECU is equivalent to $3.24 \times 10^{-7}$~\DMunit{}.
Assuming typical ionospheric magnetic field strengths of ${\sim}0.5\,$G, the equivalent change in \ac{rm} is
\begin{equation}
\Delta {\rm RM}
  = 0.132\,{\rm rad}/{\rm m}^2
     \left( \frac{\left\langle B_\parallel \right\rangle}{0.5\,{\rm G}} \right)
     \left( \frac{\Delta \rm TEC}{\rm TECU} \right),
\end{equation}
which exceeds $1\,$rad$/$m$^2$ for $\Delta {\rm TEC} = 10\,$TECU$/$min, and would be able to account for the variation.
However, the observation was taken during the quiescent part of the Solar activity cycle, which is supported by the \software{ionFR} results for the hours leading up to and immediately following the observation, over which the estimated \ac{rm} contribution remained steady.
Solar activity as measured by the F10.7\,cm emission was at a historical low of 70\,SFU\footnote{\href{https://lasp.colorado.edu/lisird/data/penticton\_radio\_flux}{https://lasp.colorado.edu/lisird/data/penticton\_radio\_flux}}, and the $K_p$ index at the time of observing was $\sim$3, indicating low levels of geomagnetic activity \citep{2021SpWea..1902641M}.
Moreover, the fact that the \ac{rm} measurements taken between 50 and 60 minutes into the observation are all above the mean value suggests that the \ac{rm} drift is more gradual, implying much larger structures in the ionosphere than those studied by \citet{{1997GeoRL..24.2283P}}.

Another possibility (to be followed up in future work) is that the RM variations are indeed intrinsic to the source, such as magnetospheric disturbances or inhomogeneities.
It is known that the magnetosphere can induce such large apparent changes in \ac{rm} at different rotation phases \citep[e.g.][]{2019MNRAS.483.2778I}, and it is possible that our sample of pulses for which we measured \acp{rm} are also sampling the magnetosphere at different locations.
In order to verify this, however, a much more careful polarimetric analysis of the individual pulses than what we have done here is warranted.

\section{Discussion}
\label{sec:discussion}

PSR \psr{} is difficult to classify.
Its position in the $P$-$\dot{P}$ diagram (Fig.~\ref{fig:PPdot}) indicates a high magnetic field (${\sim}1.5 \times 10^{13}\,$G) and average characteristic age (0.17~Myr), somewhere on the outskirts of the distribution of known \acp{rrat}.
With this age, and given its Galactic latitude, we can derive the minimum speed required for the pulsar to get to its current location assuming it was born in the Galactic plane.
For the most conservative (i.e. smallest) distance of ${\sim}0.4\,$kpc, the implied speed is $\gtrsim 2300\,$km$/$s, which is much faster than typical neutron star speeds \citep{2009PASP..121..814O}.
We argue that this is strong evidence that the spindown rate is grossly overestimated, and expect this to be resolved by more follow-up observations and a dedicated timing campaign.

\psr{}'s pulse energy distribution has an unusually long tail, making it more readily detectable via its bright, single pulses than via its periodicity.
However, it is evidently not an \ac{rrat}; as far as can be determined with the \ac{mwa}'s sensitivity, it does not null.
The established \acp{rrat} J1854+0306 \citep{10.1093/mnras/stae973} and J0628+0909 were recently found to emit much weaker pulses than had hitherto been observed, and persistent emission has also been found in several other \acp{rrat} \citep{2024MNRAS.527.4397M}.
B0826$-$34 is known to switch between pulsar-like and \ac{rrat}-like modes \citep{2012ApJ...759L...3E}, although such modes last for up to thousands of pulses at a time.
These examples raise questions about whether there is a wider population of neutron stars that bridge the gap between \acp{rrat} and normal pulsars.

The brighter pulses tend to arrive at later rotation phases than their dimmer counterparts, but it remains unclear whether there are two distinct populations.
Without a proper understanding of the dimmer pulses, which are difficult to characterize due to their low \ac{snr}, we cannot comment on whether the bright pulses should be classified as giant pulses \citep{2004IAUS..218..315J}.
The concept of a ``waiting time'' between pulses becomes ambiguous, as its value clearly depends on the choice of energy cutoff between the two populations.

It is also unclear whether the dim and bright pulses are generated by the same emission mechanism.
There may be a continuum of emission states, with the pulses transitioning towards the trailing edge of the pulse window as they grow in brightness.
One way this could be effected, for example, is by means of a radius-to-brightness mapping, analogous to the more familiar \acl{rfm} \citep{1978ApJ...222.1006C}, in which brighter pulses are emitted at higher altitudes in the pulsar's magnetosphere, making them subject to aberration and retardation effects \citep{1991ApJ...370..643B,2004ApJ...607..939D}.

On the other hand, the fact that the brightest pulses arrive at preferentially later rotation phases suggests that the two populations might in reality be distinct.
As can be seen in Fig.~\ref{fig:snr_vs_phase}, that the moderately bright pulses (with \ac{snr} in the range ${\sim}$5 to 30) occur preferentially at leading phases, giving the appearance of a distinct population from the sparser distribution on the trailing side \citep[cf. the phase dependence of the two emission modes in][]{2012ApJ...759L...3E}.
Much longer observations would be needed to see how sharply the density of these medium-brightness pulses changes as a function of pulse phase.

If the bright pulses are a kind of ``giant pulse'', then the fact that they occupy a generally different phase range isn't without precedent.
\citet{2006ApJ...640..941K} found that giant pulses from the millisecond pulsar J0218+4232 occur at unusual phases, and similar behavior is seen in other pulsars, such as B0031$-$07 \citep{2004A&A...427..575K} and the Crab pulsar \citep{Bhat_2008}.
However, the giant pulses in these cases are typically much narrower than their normal pulses, which we don't find to be the case in \psr{}.

We considered the possibility that the dimmer population is actually an interpulse (with the true period being double what we have reported in this paper), which can appear at some finite phase offset from exactly $180^\circ$ from the main pulse \citep[e.g.][]{2019MNRAS.490.4565J}.
However, in this scenario, one would expect the pulses on one side (or the other) of the pulse window to preferentially have the same remainder modulo 2, which we did not find to be the case.

Yet another possible explanation of the brightness-phase correlation is if \psr{} has a rotation phase dependent spectral index, with the trailing side having a steeper spectral index than the latter.
This was exactly the explanation put forward by \citet{2003A&A...407..655K} for the unusually bright pulses of B1133+16, which also had a preferential phase (the trailing edge of the leading component).
More sensitive observations over a wider frequency range would be needed to determine if the same is true for \psr{}.

The erratic polarization behavior of the bright single pulses is also puzzling.
The single pulses of most pulsars \citep[and many \acp{rrat}, e.g.][]{2009MNRAS.396L..95K,10.1093/mnras/stae973,2023MNRAS.518.1418H}, although they can vary, do not usually deviate from their time-averaged behavior (typically an \ac{rvm}-like curve) so much that the integrated profile becomes so completely depolarized, as appears to be the case for \psr{}.
Such variation indicates extreme turbulence at the emission site of (at least) the bright pulses.
As is apparent in Fig.~\ref{fig:RM_results}, the \ac{rm} also fluctuates significantly throughout the observation, especially between the 50 and 70 minutes marks.
As noted in \S\ref{sec:polarization}, however, we cannot rule out the possibility that this fluctuation is due to ionospheric activity.

\begin{figure}[t]
    \centering
    \includegraphics[width=\linewidth]{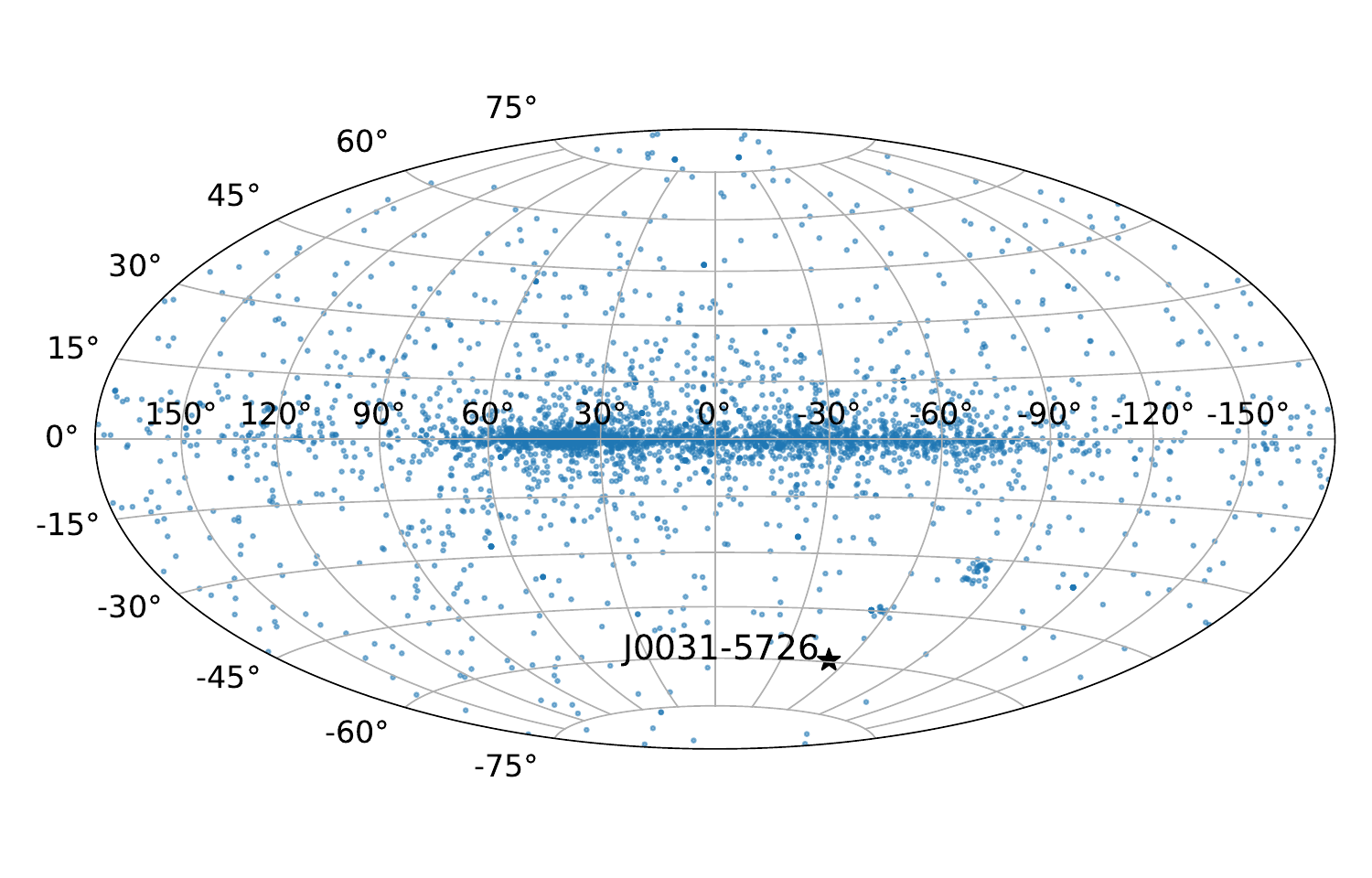}
    \caption{The Galactic position of PSR J0031-5726 (black star) amongst the known pulsar population (blue dots).}
    \label{fig:galactic_position}
\end{figure}

Regardless of its status as an \ac{rrat}, \psr{} may be useful for probing the \ac{ism} at high Galactic latitudes, as shown in Fig.~\ref{fig:galactic_position}, as it in a part of the sky relatively devoid of known pulsars.
The NE2001 \citep{2002astro.ph..7156C} and YMW16 models \citep{2017ApJ...835...29Y} models give\footnote{PyGEDM, \url{https://apps.datacentral.org.au/pygedm/} \citep{2021PASA...38...38P}} distances of $0.37\,$kpc and $0.59\,$kpc, respectively.
We estimate a ${\sim}40$\% uncertainty on these distances by comparing the accuracy of the models for 33 pulsars within a $20^\circ$ radius of \psr{} with independently determined distance measurements. Very Long Baseline Inteferometric (VLBI) parallax observations would yield an independent distance estimate, that could then constrain the electron density along the line-of-sight to this direction.

Both electron density models predict scattering timescales of order ${\sim}1$\,ns (at $1$\,GHz).
Assuming Kolmogorov turbulence, the diffractive scintillation bandwidth at the \ac{vcs} observing frequency of $154$\,MHz is expected to be a few tens of kHz \citep{1985ApJ...288..221C}.
The dynamic spectra of a few of the brightest pulses showed fluctuations that may be due to diffractive scintillation, but the excision of low \ac{snr} channels due to the response of the \ac{mwa}'s polyphase filterbank, combined with the relatively low per-channel \ac{snr}, made it difficult to make a robust measurement with these data.

\section{Conclusions}
\label{sec:conclusions}

The discovery of PSR \psr{} highlights the advantages of image-domain searches for unusual pulsars (and interesting radio transients in general) particularly with the impressive survey speed inherent in the use of wide-field instruments such as the \ac{mwa}.
In addition, the fact that voltage data is kept in the \ac{mwa} archive is what allowed us to confirm it as a pulsar and obtain preliminary timing parameters without needing extra telescope time.

\psr{} exhibits a wide range of single pulse behaviors, including a long-tailed pulse energy distribution, erratic and strongly varying \ac{pa} curves, and a tendency for the brighter pulses to occur preferentially towards the trailing edge of the pulse window.
Although many of these behaviors are seen in other pulsars and \acp{rrat}, it remains unclear whether \psr{} itself should be classified as an \ac{rrat}, or whether it is more properly described as something that helps to bridge the gap between normal pulsars and \acp{rrat}.
This is in part due to not being able to tell whether the dimmer pulses, which had very low \ac{snr} in our data set, represent a distinct form of emission from the brightest pulses.
Further studies with more sensitive telescopes may help better characterize the energy distribution of the dimmer pulses, which would shed light on this question.

\section*{Acknowledgements}

This scientific work uses data obtained from Inyarrimanha Ilgari Bundara / the Murchison Radio-astronomy Observatory. We acknowledge the Wajarri Yamaji People as the Traditional Owners and native title holders of the Observatory site. Establishment of CSIRO's Murchison Radio-astronomy Observatory is an initiative of the Australian Government, with support from the Government of Western Australia and the Science and Industry Endowment Fund. Support for the operation of the \ac{mwa} is provided by the Australian Government's National Collaborative Research Infrastructure Strategy (NCRIS), under a contract to Curtin University administered by Astronomy Australia Ltd (AAL).

NHW is supported by an Australian Research Council Future Fellowship (project number FT190100231) funded by the Australian Government.

This work was supported by resources provided by the Pawsey Supercomputing Research Centre with funding from the Australian Government and the Government of Western Australia.
This work was supported by resources awarded under AAL's ASTAC merit allocation scheme on the OzSTAR national facility (Ngarrgu Tindebeek) at Swinburne University of Technology. The OzSTAR program receives funding in part from the Astronomy NCRIS allocation provided by the Australian Government, and from the Victorian Higher Education State Investment Fund (VHESIF) provided by the Victorian Government.

SM would like to thank A. Waszewski and C. Lee for useful discussions on the interpretation of the \ac{rm} variations.

\bibliography{References}

\end{document}